\documentstyle[amssym]{mn}
 at 10truept
\input{epsf.sty}

\def\eps@scaling{1}				
\def\epsscale#1{\def\eps@scaling{#1}}
\def\epsfsize#1#2{\eps@scaling#1}

\def\kmsm{{\,{\rm km}\,{\rm s}^{-1}\,{\rm Mpc}^{-1}}}

\title
     [Observations of B1152+199]
{\vglue-3.0truecm
\vglue 2.5truecm
\noindent
High resolution observations and mass modelling of the CLASS gravitational 
lens B1152+199
\author[D. Rusin et al.]
     {D. Rusin$^1$, M. Norbury$^2$, A.D. Biggs$^2$, D.R. Marlow$^2$, 
	N.J. Jackson$^2$, 
	\newauthor
	I.W.A. Browne$^2$, P.N. Wilkinson$^2$, S.T. Myers$^3$\\ 	
	$^1$Department of Physics \& Astronomy, University of Pennsylvania,
	Philadelphia, PA 19104-6396\\
        $^2$Jodrell Bank Observatory, University of Manchester,
	Macclesfield, Cheshire, SK11 9DL\\
        $^3$National Radio Astronomy Observatory, P.O. Box 0, Socorro,
	NM, 87801}
}

\newcommand{\be}{\begin{equation}}
\newcommand{\ee}{\end{equation}}
\newcommand{\ba}{\begin{eqnarray}}
\newcommand{\ea}{\end{eqnarray}}

\newcommand{\simpropto}{\!\!\begin{array}{c} {\propto} \\
                  [-1.7ex] \sim \end{array}\!\!}

\def\spose#1{\hbox to 0pt{#1\hss}}
\def\simlt{\mathrel{\spose{\lower 3pt\hbox{$\mathchar"218$}}
     \raise 2.0pt\hbox{$\mathchar"13C$}}}
\def\simgt{\mathrel{\spose{\lower 3pt\hbox{$\mathchar"218$}}
     \raise 2.0pt\hbox{$\mathchar"13E$}}}
\def\simpropto{\mathrel{\spose{\lower 3pt\hbox{$\mathchar"218$}}
     \raise 2.0pt\hbox{$\propto$}}}

\begin{document}

\maketitle

\begin{abstract}

We present a series of high resolution radio and optical observations
of the CLASS gravitational lens system B1152+199 obtained with the
Multi-Element Radio-Linked Interferometer Network (MERLIN), Very Long
Baseline Array (VLBA) and Hubble Space Telescope (HST). Based on the
milliarcsecond-scale substructure of the lensed radio components and
precise optical astrometry for the lensing galaxy, we construct models
for the system and place constraints on the galaxy mass profile. For a
single galaxy model with surface mass density $\Sigma(r) \propto
r^{-\beta}$, we find that $0.95 \leq \beta \leq 1.21$ at $2\sigma$
confidence. Including a second deflector to represent a possible
satellite galaxy of the primary lens leads to slightly steeper mass
profiles.

\end{abstract}

\begin{keywords}
gravitational lensing -- galaxies: structure
\end{keywords}


\section{Introduction}
\label{intro}

Multiple-image gravitational lens systems are potent tools for
investigating a wide range of astrophysical and cosmological
issues. One promising application of strong lensing is the
determination of the Hubble parameter through the measurement of
differential time delays (Refsdal 1964; Schechter et al.\ 1997;
Kundi\'c et al.\ 1997; Lovell et al.\ 1998; Wisotzki et al.\ 1998;
Biggs et al.\ 1999; Fassnacht et al.\ 1999; Koopmans et al.\ 2000;
Patnaik \& Narasimha 2001). In addition, lenses can directly constrain
the inner several kiloparsecs of galaxy mass distributions (e.g.,
Kochanek 1991, 1995; Rusin \& Ma 2001; Cohn et al.\ 2001; Mu\~noz,
Kochanek \& Keeton 2001). Much of the recent interest in finding
arcsecond-scale gravitational lenses through systematic searches such
as the Cosmic Lens All-Sky Survey (e.g., Myers et al.\ 1995, 1999) and
the southern survey of Winn et al.\ (2000) has been driven by these
goals.

High resolution observations are the essential step in transforming
lens systems from mere novelties into useful tools.  For example,
optical or near-infrared imaging with the Hubble Space Telescope (HST)
is necessary to pinpoint the centre of the lensing galaxy, determine
its shape and orientation, and search for nearby objects that may be
perturbing the potential. This is complemented by observations with
high resolution radio arrays such as the Multi-Element Radio-Linked
Interferometer Network (MERLIN) and the Very Long Baseline Array
(VLBA), which offer precision astrometry and search for
milliarcsecond-scale substructure in the lensed radio
components. Taken together, these data provide the raw material for
developing models of the gravitational potential, from which the
structure of the lensing galaxy is investigated and time delays are
predicted for use in Hubble constant determination.

B1152+199 (Myers et al.\ 1999) was identified as a gravitational lens
candidate in the third phase of the CLASS survey observations. The
source consists of two compact radio components with a flux density
ratio of $\sim$ 3:1, separated by $1\farcs56$. Follow-up observations
with the Very Large Array (VLA) demonstrated that the components have
nearly identical two-point spectral indices between $8.4$ and $15$
GHz, as expected for images of a single lensed source. The lensing
hypothesis was definitively confirmed via follow-up spectroscopy
obtained with the Keck II telescope, which detected lines from a
background quasar at $z_s = 1.019$ and a foreground galaxy at $z_d =
0.439$. Imaging of the system with the Palomar 5-m telescope revealed
a bright ($m_g = 16.5$, $m_i=16.6$ in Gunn magnitudes) stellar-like
object at the radio position. The lack of structure suggested that the
fainter of the images may be attenuated by dust extinction in the
lensing galaxy. Subsequent observations with the Nordic Optical
Telescope (Toft, Hjorth \& Burud 2000) detected a weak counterpart to
the secondary lensed image. The large optical flux ratio between the
images ($\simeq$ 60:1 at V band) points to significant differential
extinction by the lensing galaxy, which was detected following the
subtraction of scaled point-spread functions (PSFs).

This paper presents vastly improved radio and optical observations of
B1152+199, and constructs models for the lensing mass distribution
based on the high resolution data.  Section 2 analyses the radio
properties of B1152+199, including expanded VLA radio spectra and deep
observations with MERLIN and the VLBA. Section 3 presents optical
imaging obtained with HST. In section 4 we investigate models for the
system and constrain the lensing galaxy mass profile. Section 5
summarises our findings and outlines additional work.

\section{Radio Imaging}
\label{radiodata}

B1152+199 consists of two radio components with a flux density ratio
of $\sim$ 3:1 and a separation of $1\farcs56$ (Myers et al.\
1999). Each of the components is unresolved by the VLA. To investigate
the spectral properties of the components over a wide range of
frequencies, the system was observed at 1.4, 5, 8.4 and 15 GHz using
the VLA in A configuration on 1999 July 15. The phase calibrator was
J1150+242, and observations of 3C286 were used to set the flux density
scale. The VLA data sets were calibrated in the Astronomical Image
Processing System (AIPS) using the standard procedure and analysed in
DIFMAP (Shepherd 1997). In each case the visibility data were fitted
to a pair of compact Gaussian components using several iterations of
model-fitting and phase-only self-calibration, with a solution
interval of 0.5 min. The component flux densities are given in Table
1, and the radio spectra are plotted in Figure~1. The spectra are
nearly identical, as expected for lensed images of a single background
source. The overall spectral indices between 1.4 and 15 GHz are
$\alpha_{1.4}^{15} = -0.07 \pm 0.01$ (A) and $\alpha_{1.4}^{15} =
-0.07 \pm 0.01$ (B), respectively. The flatness of the radio spectra
suggests that the source may exhibit sufficient variability to allow
for the measurement of a differential time delay. However, trial
monitoring observations using the VLA at 8.4 GHz have thus far failed
to detect any significant variability ($\la 2\%$) in the lensed
components.

MERLIN 5 GHz observations of B1152+199 were performed on 1999 January
2 for 7 hr and again on 1999 January 5 for 14 hr. The flux density
scale was determined by 15 min observations of 3C286 and the point
source calibrator OQ208. Alternate observations of the target source
(8 min) and a nearby phase reference source J1148+186 (2 min) were
iterated. The combined data were calibrated in AIPS and imaged in
DIFMAP by repeating a cycle of cleaning and phase-only
self-calibration, starting with long solution intervals ($40$ min) and
gradually decreasing to a minimum interval of $2$ min. Once the model
had sufficiently converged, an amplitude self-calibration was
performed using a solution interval of 30 min.  The final map has an
rms noise of $70$ $\mu$Jy/beam and is shown in Figure~2. The data were
modelled by two compact Gaussian components with flux densities of
$52.6 \pm 0.1$ mJy (A) and $18.2 \pm 0.1$ mJy (B). No further emission
was detected down to the $3\sigma$ level of the residual map.

\begin{figure}
\begin{center}
\leavevmode
\epsscale{.75}
\centerline{
\epsfbox{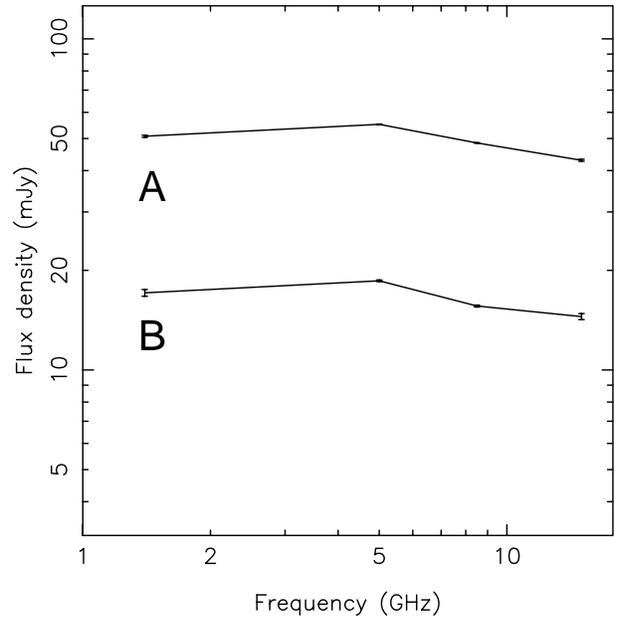}
}
\caption[1]{B1152+199 component radio spectra based on the VLA 1.4, 5,
8.4 and 15 GHz data of 1999 July 15.}
\end{center}
\end{figure}

\begin{figure}
\begin{center}
\leavevmode
\epsscale{.4}
\centerline{
\epsfbox{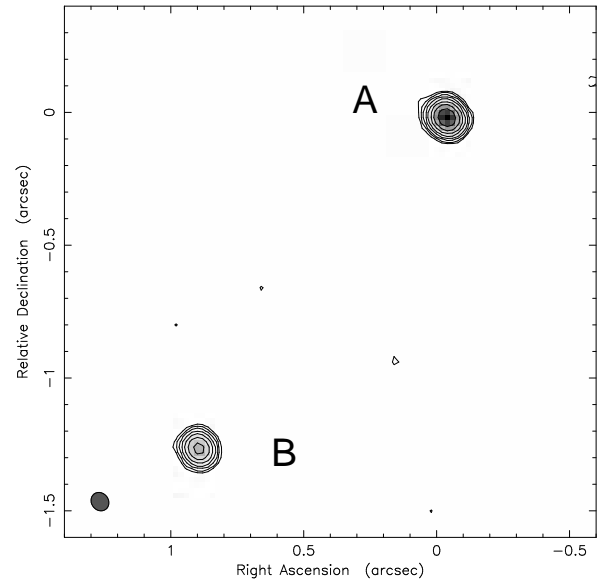}
}
\caption[2]{MERLIN 5 GHz observation of B1152+199 taken 1999 January 2 and 5.
The lowest contour is $3\times$ the map rms noise of $70$ $\mu$Jy/beam, and
contour levels increase by factors of two. The restoring beam is $72.5\times
63.7$ mas at $+38.1^{\circ}$. The data have been naturally weighted.}
\end{center}
\end{figure}

\begin{table}
\caption{B1152+199 component flux densities (in mJy) at 1.4, 5, 8.4
and 15 GHz. Data for the 1999 July 15 VLA observation. Errors in the
flux densities are taken to be equal to the rms noise of the
respective maps.}
\begin{tabular}{c c c c c}
Comp & $S_{1.4}$ & $S_{5}$ & $S_{8.4}$ & $S_{15}$ \\
A & $50.8 \pm 0.4$ & $55.2 \pm 0.1$ & $48.5 \pm 0.1$ & $43.0 \pm 0.3$\\
B & $17.1 \pm 0.4$ & $18.6 \pm 0.1$ & $15.6 \pm 0.1$ & $14.5 \pm 0.3$\\
\end{tabular}
\end{table}

\begin{figure*}
\epsscale{1.2}
\centerline{
\epsfbox{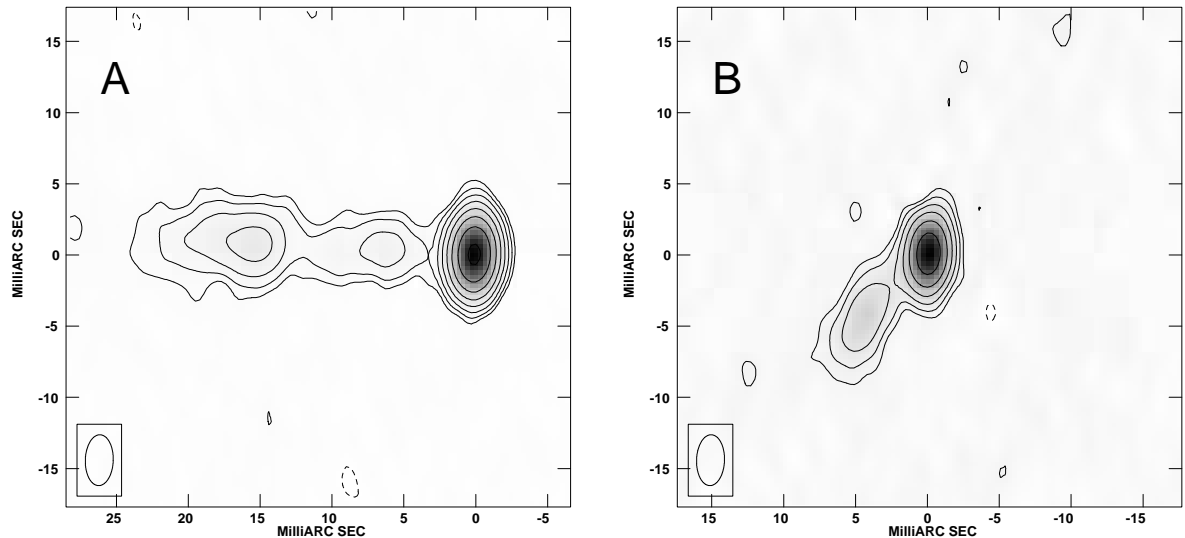}
}
\caption{VLBA 5 GHz observation of B1152+199 taken 2001 February 27 and
March 18. The lowest contour is $3\times$ the map rms noise of $75$
$\mu$Jy/beam, and contour levels increase by factors of two. The beam is 
$3.6 \times 1.9$ mas at $-1.8^{\circ}$. The data have been naturally
weighted. Left: Component A. Right: Component B.}
\end{figure*}

\begin{table}
\caption{B1152+199 component substructure. Data from the VLBA 5 GHz
observation of 2001 February 27 and 2001 March 18. Listed are the
coordinates of each model component, its flux density and model-fitted
position angle (for the Gaussian jets).  Errors on the positions are
$\simeq 0.1$ mas. All errors are the nominal $1\sigma$ uncertainties
estimated by the AIPS OMFIT task.}
\begin{tabular}{l c c r r}
Comp & $x_1$ & $x_2$ & $S_{5}$ (mJy) & $PA_{Gauss}$\\
$\rm A_{core}$ & $\equiv 0$ & $\equiv 0$ & $33.4 \pm 0.2$ \\
$\rm A_{ jet}$ &$+0\farcs0115$ & $+0\farcs0006$ & $17.1 \pm 0.1$ & $+87^{\circ}
\pm 1^{\circ}$\\
$\rm B_{core}$ &$+0\farcs9353$ & $-1\farcs2454$ & $11.9 \pm 0.1$\\
$\rm B_{ jet}$ &$+0\farcs9394$ & $-1\farcs2494$ & $5.0 \pm 0.1$ & $+143^{\circ}
\pm 2^{\circ}$\\
\end{tabular}
\end{table}

VLBA 5 GHz observations of B1152+199 were obtained on 2001 February 27
and March 18. The observing time was three hours per epoch, which was
divided among three widely spaced hour angles to maximise the $uv$
coverage. The data were recorded in two IFs, each of which was split
into 16 0.5-MHz channels giving a total bandwidth of 16~MHz. The data
were digitized using two-bit sampling and both the Stokes L and R
polarizations were recorded.  The integration time per visibility was
2 seconds. A five minute scan of the bright source 4C39.25 was
included for fringe-finding purposes.  Alternate observations of the
target source (3 min) and the nearby phase reference source J1148+186
(1.5 min) were then iterated.

Data reduction was carried out in AIPS on the combined epochs. The
data were first flagged and amplitude calibrated using recorded system
temperatures and known telescope gain curves.  Next the data were
fringe-fitted, using 4C39.25 to derive an initial delay
correction. The phase calibrator was then used to find the residual
rate, phase and delay solutions as a function of time. This was
initially fringe-fitted assuming that it was a point source and later
mapped and self-calibrated to produce a model that included the
effects of resolved structure. The solutions so found were then
applied to B1152+199. The data were mapped using the IMAGR task. As
the correlation centre of the data was a position close (50 mas) to
component A, the data were not averaged in time or frequency so as to
avoid any smearing in the final images. Maps were made of two fields
at the positions of the lensed radio components. Several iterations of
mapping and phase-only self-calibration were performed until the CLEAN
model converged. The data were then amplitude self-calibrated using a
long solution interval (80 min). The resulting maps are displayed in
Figure~3 and have an rms noise of $75 \mu$Jy/beam. Each of the
components is resolved into a compact core and extended jet
structure. Finally, the data were model-fitted using the OMFIT
task. For each image, two components were used to crudely describe the
observed structure: a delta function for the core and a Gaussian for
the jet. The positions, flux densities and orientations of these
components are listed in Table~2.

Note the apparent peculiarity of the B1152+192 radio substructure. In
component A the jet is quite straight. Specifically, the position
angle of the Gaussian jet component ($+87^{\circ}$) is nearly
identical to that of an axis drawn between the centre of that
component and the core. The jet in B, however, appears to be slightly
bent.  Here the position angle of the Gaussian jet ($+143^{\circ}$) is
significantly larger than that of an axis passing through the centres
of the core and jet model components ($+134^{\circ}$). The bending of
B could be produced by substructure in the lensing mass distribution
(e.g., Metcalf \& Madau 2001). Deeper imaging at higher resolution
will be essential for confirming this discrepancy and evaluating
possible solutions.

\section{Hubble Space Telescope Imaging}
\label{HST}

\begin{figure}
\begin{center}
\leavevmode
\epsscale{.45}
\centerline{
\epsfbox{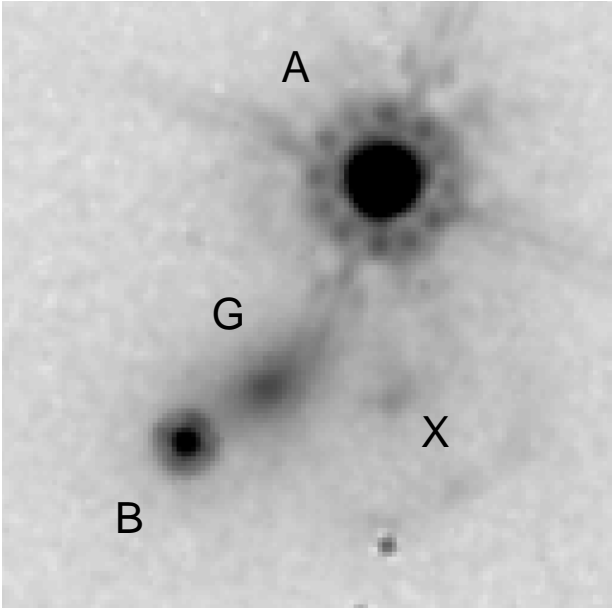}
}
\end{center}
\caption{HST WFPC2 I-band image obtained 2000 March 28. The shading is
logarithmic. Counterparts to the lensed quasar images (A \& B), the
lensing galaxy (G) and a possible satellite galaxy (X) are
detected. The region displayed is $3'' \times 3''$.}
\end{figure}

HST observations of B1152+199 were obtained on 2000 March 28 with the
Wide Field and Planetary Camera 2 (WFPC2). The F555W (``V-band'') and
F814W (``I-band'') filters were used. For each filter, four dithered
exposures of 500 sec duration were obtained. A standard reduction was
performed on the data sets using the Image Reduction and Analysis
Facility (IRAF). The resulting I-band image is displayed in Figure
4. The V-band image detected all the same features but at a lower
signal-to-noise, and is not shown. The lensing galaxy (G) and
counterparts to the lensed quasar components (A $\&$ B) are clearly
detected. In addition, a faint extended emission feature (X) is seen
to the west of G. Component X certainly cannot be an additional lensed
image, as it has no radio counterpart. It may therefore be a satellite
galaxy of the primary lens. In addition, evidence of a diffuse arc is
seen to the west of image B. This may be lensed extended emission from
the quasar host galaxy. Its position would require that part of the
host galaxy resides in the region enclosed by the tangential caustic
curve, inside of which sources are lensed into four images.

\begin{table}
\caption{HST photometry and astrometry. Listed are the positions of the
components derived from the I-band image, and their integrated magnitudes
$m_V$ and $m_I$. The magnitude of X was calculated using only the central
$150\times 150$ mas. Errors are $10$ mas on the positions and $\simeq 0.3$ on
the magnitudes.}
\begin{tabular}{c c c r r}
Comp & $x_1$ & $x_2$ & $m_V$ & $m_I$\\ 
A & $\equiv 0$    & $\equiv 0$ & $\la 17.3$ & $\la 16.7$\\
B & $+0\farcs942$ & $-1\farcs251$ 	  & $21.9$  & $19.9$\\
G & $+0\farcs554$ & $-0\farcs991$ & $22.6$  & $19.6$\\ 
X & $-0\farcs037$ & $-1\farcs046$ & $25.0$  & $23.1$\\
\end{tabular}
\end{table}

Photometry and relative astrometry were performed on the HST data. The
positions and magnitudes of the components are listed in Table
3. There were difficulties in subtracting PSFs because image A is
probably saturated. Consequently, the errors in the magnitudes are
quite large ($\simeq 0.3$). The relative optical and radio positions
of A and B are consistent to within the HST error bars ($\simeq 10$
mas). The position of G differs from the ground-based measurement of
Toft, Hjorth \& Burud (2000) by more than $100$ mas, which
demonstrates the importance of HST observations for investigating lens
systems. Finally, elliptical isophotes were fit to G. At a distance of
$225$ mas from the galaxy centre we find a surface brightness axial
ratio of $f = 0.84 \pm 0.9$ and a position angle of $-63^{\circ} \pm
17^{\circ}$. Closer to the galaxy centre, the isophotes tend to have
smaller axial ratios and somewhat more vertical orientations.

\section{Mass Modelling}

We now investigate simple models for the B1152+199 lensing
potential. The models were constrained using the positions ($x_1$,
$x_2$) of the two lensed core components from the VLBA data and the
flux density ratio. The consistency of the measured flux density ratio
among the various radio data sets strongly suggests that the value has
been affected very little by variability and is therefore close to the
true magnification ratio. We thus set $r=|S_A/S_B| = 3.0$ and assumed
an uncertainty of $\Delta r = 0.15$ (5\%) to account for possible
effects from low-level variability or model-fitting errors.  We set
$\Delta x_{1} = \Delta x_2 = 0.1$ mas, and the lens and source
redshifts to their measured values of 0.439 and 1.019, respectively
(Myers et al.\ 1999). A flat $\Omega_{\Lambda}=0.7$ cosmology with
$H_{0} = 100h \kmsm$ was assumed for all calculations.

Because the lensing mass in B1152+199 is dominated by a single galaxy, we
modelled the system using a singular power-law ellipsoid (SPLE) mass
distribution (e.g., Barkana 1998), with scaled surface density
\begin{eqnarray}
\kappa(x_1,x_2) = \frac{q}{ (x_1^2 + x_2^2 / f^2)^{\beta/2}} 
\end{eqnarray}
where $\beta$ is the power-law slope, $q$ is the normalisation and $f$
is the projected axial ratio. Current constraints on the inner several
kiloparcsecs of elliptical galaxies (e.g., Cohn et al.\ 2001; Grogin
\& Narayan 1996; Kochanek 1995; Koopmans \& Fassnacht 1999; Rix et
al.\ 1997; Rusin \& Ma 2001) favour mass distributions that are close
to isothermal ($\beta = 1$), but a range of slopes ($0.8 \leq \beta
\leq 1.2$) is also consistent with the data. We considered various
values of $\beta$ in this regime. Our calculations made use of the
rapidly converging series solutions for the deflection angles and
magnification matrices of power-law mass profiles derived by Barkana
(1998) and implemented in the ``FASTELL'' software package.

Lens modelling was performed with an image plane minimisation (Kochanek 1991),
which optimised the fit statistic
\begin{equation}
\chi^{2} = \sum_{i=A,B} \left[ \frac { (x_{i,1}' - x_{i,1})^{2}}{\Delta
x_{i,1}^{2}}  +  \frac { (x_{i,2}' - x_{i,2})^{2}}{\Delta
x_{i,2}^{2}}     \right] +
\frac{(r' -r )^{2}}{\Delta r^{2}}
\end{equation}
where primed quantities are model-predicted and unprimed quantities
are observed. To reduce the number of free parameters and ensure a
constrained model, the position of the galaxy was fixed at
($+0\farcs544$, $-0\farcs991$) relative to A. With the profile slope
also fixed, the model has five parameters: the normalisation, axial
ratio and position angle of the SPLE, and two coordinates for the
unlensed source. The number of degrees of freedom (NDF) is zero, so
the best-fitting model should be described by $\chi^2 = 0$, regardless
of the assumed uncertainties. Monte Carlo simulations were performed
to examine the stability of the models. Gaussian-distributed errors
accounting for observational uncertainties were added simultaneously
to the galaxy position ($10$ mas in each coordinate) and flux density
ratio ($5\%$). The models were optimised for each of 10000 trials. The
uncertainties in the parameters were found from the ranges enclosing
$95\%$ of the results.

The optimised model parameters and time delays for various values of
$\beta$ are listed in Table 4. The differential time delay for the
isothermal ($\beta = 1$) model is $35.9 \pm 2.0$ $h^{-1} $ days in a
flat $\Omega_{\Lambda} = 0.7$ cosmology. The delay varies linearly
with the assumed profile slope, as predicted by Witt, Mao \& Keeton
(2001). Critical curves and caustics for the isothermal case are
plotted in Figure~5.

\begin{table*}
 \centering \begin{minipage}{140mm} \caption{Lens modelling results
 for power-law deflectors with different mass profiles $\beta$. Listed
 are the axis ratio $f$, galaxy position angle $PA_G$ (in degrees),
 predicted time delay $\Delta t$ (in $h^{-1}$ days), magnification
 matrix elements $A_A$ and $A_B$ at the image positions, and the
 predicted position angle (in degrees) of component B assuming that
 $PA_A = +87^\circ$. Uncertainties are determined from the range
 enclosing 95\% of the Monte Carlo results. All values assume a flat
 $\Omega_{\Lambda} = 0.7$ cosmology, and were calculated using the fit
 parameter defined in eq. (2).}

\begin{tabular}{cccccc}
 & $\beta = 0.8$ & $\beta = 0.9$ & $\beta = 1.0$ & $\beta = 1.1$ &
 $\beta = 1.2$ \\ & \\ $f$ & $0.787_{-0.025}^{+0.024}$ &
 $0.755_{-0.030}^{+0.029}$ & $0.719_{-0.035}^{+0.034}$ &
 $0.677_{-0.041}^{+0.040}$ & $0.627_{-0.048}^{+0.046}$ \\


$PA_G$ &  
$-73.4_{-3.6}^{+3.3}$ &			
$-76.3_{-4.0}^{+3.6}$ &
$-79.4_{-4.2}^{+3.9}$ &
$-82.6_{-4.4}^{+4.2}$ & 
$-85.6_{-4.4}^{+4.3}$ \\

$\Delta t$ &  
$28.3 \pm 1.6$ &
$32.1 \pm 1.8$ &
$35.9 \pm 2.0$ &
$39.7_{-2.2}^{+2.3}$ &
$43.4_{-2.4}^{+2.5}$ \\

$A_{A,11}$ 	&  
$+0.379_{-0.017}^{+0.018}$ &		
$+0.430_{-0.020}^{+0.021}$ &
$+0.482_{-0.022}^{+0.024}$ &
$+0.533_{-0.025}^{+0.026}$ &
$+0.582_{-0.026}^{+0.027}$ \\

$A_{A,12}$ 	&  
$-0.273_{-0.010}^{+0.011}$ &
$-0.284 \pm 0.013$ & 
$-0.290 \pm 0.015$ &
$-0.292 \pm 0.017$ & 
$-0.292 \pm 0.018$ \\

$A_{A,22}$ 	&  
$+0.682_{-0.010}^{+0.009}$&
$+0.762_{-0.012}^{+0.011}$&
$+0.838 \pm 0.013$&
$+0.910_{-0.015}^{+0.014}$&
$+0.975_{-0.016}^{+0.015}$ \\

$A_{B,11}$ 	&  
$+0.247_{-0.056}^{+0.054}$&		
$+0.317_{-0.062}^{+0.060}$&
$+0.396_{-0.067}^{+0.066}$&
$+0.484 \pm 0.073$&
$+0.580_{-0.079}^{+0.078}$ \\

$A_{B,12}$ 	&  
$-0.681_{-0.027}^{+0.028}$&		
$-0.792_{-0.032}^{+0.033}$&
$-0.905_{-0.037}^{+0.038}$&
$-1.021_{-0.041}^{+0.042}$&
$-1.137_{-0.045}^{+0.046}$ \\

$A_{B,22}$ 	&  
$-0.356_{-0.103}^{+0.097}$&		
$-0.366_{-0.116}^{+0.109}$&
$-0.357_{-0.127}^{+0.119}$&
$-0.324_{-0.134}^{+0.126}$&
$-0.265_{-0.138}^{+0.131}$ \\

$PA_B$		&  
$+123.0_{-3.9}^{+4.1}$&		
$+126.1_{-4.0}^{+4.3}$&
$+129.9_{-4.2}^{+4.5}$&
$+134.5_{-4.4}^{+4.8}$&
$+140.0_{-4.7}^{+5.1}$ \\

\end{tabular}
\end{minipage}
\end{table*}

\begin{figure}
\begin{center}
\leavevmode
\epsscale{.75}
\centerline{
\epsfbox{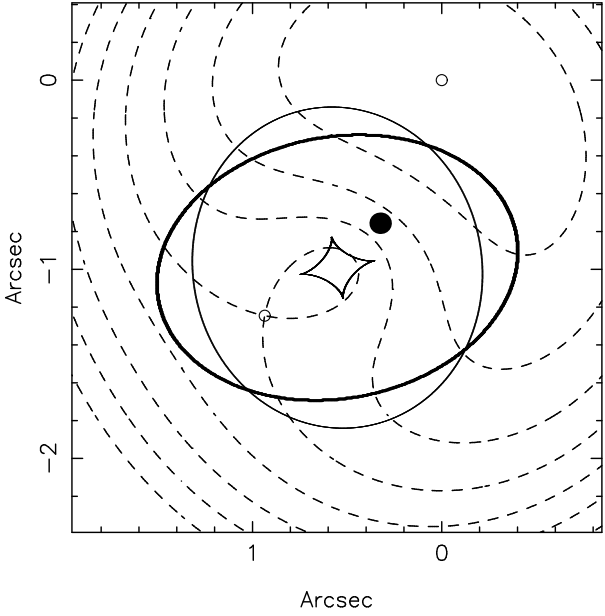}
}
\end{center}
\caption{The critical curve (thick line) and caustics (thin lines) of the
isothermal ($\beta = 1$) lens model. The filled circle marks the recovered
source position.  The open circles indicate the positions of the images.
Dashed lines denote contours of constant time delay in increments of $9.0
h^{-1}$ days outward from the global minimum at image A.}
\end{figure}

The above models can be subjected to a pair of consistency
checks. First, the mass distributions of ellipticals tend to be
well-aligned with the light (Keeton, Kochanek \& Falco 1998; Kochanek
2001), so it is useful to compare the position angles of the SPLEs
with the surface brightness of the lensing galaxy. From Table 4 we see
that the model-predicted position angles are generally compatible with
the HST value, but the agreement diminishes as the profile is made
steeper. The discrepancies are, however, broadly consistent with the
range found by Kochanek (2001). Offsets between the mass and light
position angles are often due to external shear fields, but we lack a
sufficient number of constraints to unravel the distribution of
internal and external shear in B1152+199 at this time.

Second, we considered the relative orientations of the B1152+199 radio
components. An infinitesimal vector $d\vec{x}_A$ associated with image
A is mapped to a vector $d\vec{x}_B$ associated with image B according
to
\begin{equation}
d\vec{x}_B = A_B^{-1}  A_A d\vec{x}_A
\end{equation}
where $A_{A,B} = \partial \vec{y} / \partial \vec{x}$ are the
magnification matrices (e.g., Schneider, Ehlers \& Falco 1992) at the
image positions A and B, respectively, and $\vec{y}$ represents a
vector in the source plane. We define the position angles of the radio
components using the axes passing through the centres of their
respective core and jet subcomponents, as our experience indicates
that this will be the cleanest indicator of orientation. Given the
values in Table 2, $PA_A = +87^{\circ} \pm 1^{\circ}$ and $PA_B =
+134^{\circ} \pm 2^{\circ}$. Using eq. (4) and the recovered
magnification matrices, we projected a vector $d\vec{x}_A$ with $PA_A
= +87^{\circ}$ to a vector $d\vec{x}_B$ and calculated its position
angle. The resulting values are listed in Table 4, with the
uncertainties determined by the Monte Carlo. Note that steeper mass
profiles lead to larger position angles for component B. A mass
profile with $\beta \simeq 1.1$ provides the best fit to the observed
orientations. The popular isothermal model is still compatible with
the VLBA data, but shallower mass profiles fail to reproduce the
substructure.

To better quantify constraints on the mass profile, we calculated a new fit
parameter 
\begin{equation}
\chi^2 = \chi^2_{img,pos} + \chi^2_{img,flx} + \chi^2_{gal,pos} +
\chi^2_{PA_B} \, ,
\end{equation}
that includes contributions from the position angle of component B
($\chi^2_{PA_B}$) and the galaxy centre ($\chi^2_{gal,pos}$), which
was now allowed to vary in order to best reproduce the
substructure. The galaxy position was constrained by the HST
astrometry, with a fit tolerance of $10$ mas on each coordinate. The
model-predicted position angle of B was again computed by fixing A at
the observed value of $+87^{\circ}$. A fit tolerance of $2^{\circ}$
was assumed for the orientation of B. The models have NDF = 1. A plot
of $\Delta \chi^2$ as a function of $\beta$ is presented in Figure
6. The value of $\Delta \chi^2$ is dominated by the contribution from
$PA_B$. The data favour a lensing mass profile with $0.95 \leq \beta
\leq 1.21$ at $2\sigma$ confidence ($\Delta \chi^2 \leq 4$). The
limits are changed only slightly when the rather arbitrary error
assigned to the flux density ratio is set to 1\% or 10\%. Finally, the
absence of additional components in the 5 GHz MERLIN map requires that
any third lensed image have a flux density $S_C$ such that $S_A / S_C
\geq 150$ (see also Rusin \& Ma 2001), where $S_A$ is the flux density
of image A, and a conservative detection limit of 5 times the map rms
noise level has been assumed. This condition is satisfied for all SPLE
mass models with $\beta \geq 0.60$. Therefore, the lack of a
detectable third image leads to no improvement in the profile
constraints in this particular system, but is consistent with the
bounds derived using the observed images.

\begin{figure}
\begin{center}
\leavevmode
\epsscale{.75}
\centerline{
\epsfbox{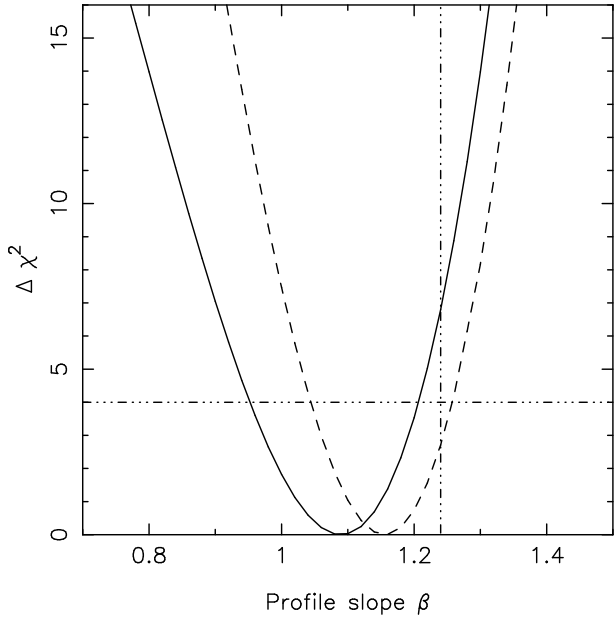}
}
\end{center}
\caption{Constraints on the lensing galaxy mass profile. Plotted is
$\Delta \chi^2$ as a function of profile slope $\beta$ for the single
deflector model (solid line) and two deflector (dashed line) model
with $q_G/q_X = 4$. The horizontal marker indicates the $2\sigma$
confidence limit. The vertical marker denotes an additional upper
bound for the two deflector model from the absence of a third 
image.}
\end{figure}

We also investigated the expected perturbation due to the possible
satellite galaxy X. Relative velocity dispersions can be estimated
from the galaxy luminosities $L$ and Faber-Jackson (1976) relation
$\sigma \propto L^{1/4}$. Because $q \propto \sigma^2 \propto
L^{1/2}$, $q_G / q_X \simeq 3$ using the V-band data and $q_G / q_X
\simeq 5$ using the I-band data.\footnote{Note that these values
assume spherical deflectors. Modifications would be required in the
presence of ellipticity (e.g., Keeton, Kochanek \& Seljak 1997). For
this preliminary analysis, however, we ignore such issues.} We fixed
the position of X according to the HST astrometry and modelled it as a
spherical deflector with the same power-law slope as G. We ran trials
with $q_G / q_X = 3$, 4 and 5, and the results are listed in Table~5.
The inclusion of X has a negligible effect on the position angle and
axial ratio of G for each $\beta$, but the time delay is decreased by
$15-25\%$ relative to the single deflector case. In addition, the
predicted value of $PA_B$ is smaller for a given $\beta$, making
steeper mass profiles necessary to reproduce the data in the two
deflector model (Table~5). Fixing $q_G / q_X = 4$, for example,
implies that $1.04 \leq \beta \leq 1.26$ at $2\sigma$ confidence. For
profiles steeper than isothermal, a third, negative-parity image can
be formed near X. This image becomes brighter as $\beta$ is increased
over the range of profile slopes investigated. The absence of the
image tightens the upper bound on the profile for more massive
secondary deflectors: $\beta \leq 1.13$ for $q_G/q_X = 3$ and $\beta
\leq 1.24$ for $q_G/q_X = 4$.

\begin{table}
\caption{Lens modelling results for two deflector mass models. Listed
are the fixed normalisation ratio $q_G/q_X$, the $2\sigma$ confidence
limit on the profile slope ($\beta$ bound I) from the observed images,
an additional upper bound from the absence of a detectable third image
($\beta$ bound II), and the time delay for the isothermal ($\beta =
1$) case in $h^{-1}$ days.}
\begin{tabular}{c c c c c}
$q_G/q_X$ & $\beta$ bound I & $\beta$ bound II & $\Delta t$\\
3 & $1.06 \leq \beta \leq 1.26$ & $\beta \leq 1.13$ & 27.7\\ 
4 & $1.04 \leq \beta \leq 1.26$ & $\beta \leq 1.24$ & 29.8\\
5 & $1.03 \leq \beta \leq 1.25$ & $\beta \leq 1.32$ & 30.8\\
\end{tabular}
\end{table}

Finally, note that if we had used a larger position angle for B,
steeper mass profiles would be required to fit the data.  Assuming
that $PA_B = +143^{\circ}$, as derived from the orientation of the
Gaussian model component in Table 2, the profile constraints become
$1.14 \leq \beta \leq 1.34$ ($2\sigma$) for the single deflector case
and $1.2 \leq \beta \leq 1.4$ ($2\sigma$) for the two deflector case
with $q_G / q_X = 3-5$. The absence of a detectable third image is
satisfied for the same range of profile slopes given in Table~5. This
rules out virtually all $\beta$ for $q_G / q_X = 3$ and 4.

\section{Discussion}

We have presented high resolution observations of the gravitational
lens CLASS B1152+199 obtained with MERLIN, the VLBA and HST. Such data
are vital if lens systems are to fulfill their potential for
constraining galaxy mass distributions or determining the Hubble
parameter. While previous spectroscopic (Myers et al. 1999) and
ground-based optical (Toft, Hjorth \& Burud 2000) observations were
sufficient to confirm B1152+199 as a gravitational lens, they offered
little detailed knowledge about the structure of the system. The
motivation for this paper was therefore to provide the high resolution
data necessary for more advanced astrophysical and cosmological
investigations of this lens.

VLA and MERLIN observations of B1152+199 show two compact radio
components separated by $1\farcs56$, with a flux density ratio of
$\sim$ 3:1. Both components have nearly identical flat radio spectra
between 1.4 and 15 GHz, as expected for lensed images of a single
background source. VLBA 5 GHz observations resolved each of the
components into a compact core and extended jet. HST observations with
the WFPC2 camera detected optical counterparts to each of the quasar
images. The flux density ratio is $\simeq 70$:1 at V band and $\simeq
20$:1 at I band, confirming the strong extinction of component B as it
passes through the lens. In addition to clearly detecting the primary
lensing mass (G), the images show a faint nearby emission (X) feature
that may be a satellite galaxy.

We investigated mass models for B1152+199 based on the new radio and
optical data. The predicted time delay for an ellipsoidal mass
distribution with an isothermal profile is $35.9 \pm 2.0$ $h^{-1}$
days, assuming a flat $\Omega_{\Lambda} = 0.7$ cosmology. The delay
varies linearly with the assumed mass profile, as predicted by Witt,
Mao \& Keeton (2001). The inclusion of a second power-law deflector to
represent the possible satellite galaxy X decreases the time delay by
$15-25\%$. The relative orientations of the lensed radio components
were shown to be powerful probes of the mass profile. Assuming a
surface mass distribution $\Sigma(r) \propto r^{-\beta}$, the
milliarcsecond-scale substructure requires that $0.95 \leq \beta \leq
1.21$ ($2 \sigma$) for a single galaxy model, and slightly steeper
profiles when X is included.  The lower limit on $\beta$ is
particularly secure, as an alternate definition of the component
orientations, based on the model-fitted position angles of the
Gaussian jets, would lead to steeper mass profiles. B1152+199 is the
simplest gravitational lens system in which both upper and lower
bounds on the lensing mass profile have been derived. This system
demonstrates the power of VLBI data for the investigation of mass
models.

Additional modelling insights are likely to be obtained from higher
resolution VLBI observations of B1152+199. Note that the jet
associated with image A appears to have a pair of hotspots (Figure
3). If corresponding features can be resolved in component B, it may
be possible to construct models based on the detailed substructure of
the jets, similar to the analysis performed on QSO 0957+561 (Barkana
et al.\ 1999). Such observations should also determine whether the
slight bend in the B jet suggested by the current VLBA 5 GHz data is
real. Observations with the HST Near Infrared Camera and Multi-Object
Spectrometer (NICMOS) will also be undertaken in an attempt to detect
and map emission from the quasar host galaxy. The information provided
by lensed extended emission is a vital ingredient for properly
reconstructing the mass distributions of lenses (Kochanek, Keeton \&
McLeod 2001). Finally, future studies will focus on measuring the
differential time delay. The system is currently being monitored at
the VLA along with several other new CLASS gravitational lens systems.

\section*{Acknowledgements}

We thank the MERLIN, VLBA and HST staffs for their assistance during our
observing runs. We also thank the referee, Alok Patnaik, for insightful
comments that helped us improve our initial draft. The National Radio
Astronomy Observatory is a facility of the National Science Foundation
operated under cooperative agreement by Associated Universities, Inc. MERLIN
is operated as a National Facility by the University of Manchester, on behalf
of the UK Particle Physics \& Astronomy Research Council. This research used
observations with the Hubble Space Telescope, obtained at the Space Telescope
Science Institute, which is operated by Associated Universities for Research
in Astronomy Inc. under NASA contract NAS5-26555. D.R. acknowledges funding
from the Zaccheus Daniel Foundation. This work was supported in part by
European Commission TMR Programme, Research Network Contract ERBFMRXCT96-0034
(``CERES'').

\end{document}